\documentclass[aps,preprint,showpacs]{revtex4}
\begin{document}        %
\draft
\title{Note on "Dynamics of Quantum Vorticity in a Random Potential"}
\author{Chu Z. K. Hua} %\cite{}} %\thanks{Present Address : }}  %address%\cite{PKU:1999}  %date
\affiliation{4/F, 16, Lane 21, Road Guang-Hui,
  Taipei 116, Taiwan, China}
%%\renewcommand{\baselinestretch}{1}  %%\renewcommand{\baselinestretch}{1.5}
%\begin{document}           % End of preamble and beginning of text.
%%\renewcommand{\baselinestretch}{1}   %%\nopagebreak
\begin{abstract}
We make comments on  Link's [{\it Phys. Rev. Lett.} 102, 131101
(2009)] paper. It seems to us that the role of vortex core was
neglected by Link during his calculation. However, the vortex core
is crucial to the interactions between the vortex and random
lattices.
%\noindent Keywords : Inflationary universe, brane inflation,
%string theory, anthropic principle .
%
\end{abstract}
\pacs{97.60.Jd, 26.60.-c, 47.37.+q, 97.60.Gb}
% 03.75.Mn, 05.30.Jp} %62.65.+k, 71.23.An, 91.60.Lj, 47.56.+r}
%03.75.Hh, 03.75.Kk, 05.30.Jp} %03.75.Hh, 03.75.Lm, 05.70.-a}
%%\end{titlepage}      %%\twocolumn     %%\nopagebreak   %\oddsidemargin=1mm
%\doublerulesep=6mm    %
%\baselineskip=6mm
\maketitle
\bibliographystyle{plain}
Link just investigated the dynamics of a superfluid vortex in a
random potential, as in the inner crust of a neutron star [1]. He
claimed that : A correct description of vortex motion in the inner
crust requires the inclusion of two additional forces: (1) the
local, nondissipative component of the force exerted on the vortex
by the lattice and (2) the elastic force of the vortex [1].
%\newline
However, as the present author checked, the vortex self-energy
$T_v$ (tension, cf. Eq. (4) in [1]) for an excitation of wave
number $k$ is not correctly used in [1] (we shall describe the
details below considering the vortex-core regime as well as either
pinning energy per unit length and  pinning energy per pinning
site [2]).
%----------
%\begin{displaymath}
% {\bf F}_{mf}=\beta' \rho_s {\bf \omega} \times ({\mathbf v}_s -{\bf
% v}_m)+\beta \rho_s {\bf \nu} \times [{\bf \omega}\times
% ({\bf v}_s -{\bf
% v}_m)],
%\end{displaymath}
%where ${\mathbf \omega}\equiv \nabla \times {\bf v}_s$ is the
%superfluid vorticity, $\nu$ is the unit vector in the direction of
%${\mathbf \omega}$.
%----------  Beltrami  state
\newline
The other remark is about the derivation of Eq. (4). As mentioned
in [1] : The force per unit length exerted on the vortex by the
lattice has a nondissipative contribution $f_0$ and a dissipative
contribution taken here to be the drag force of Eq. (2) with
$\eta'=0$, assumed to hold locally, and approximated as linear in
the local vortex velocity (but easily generalized). The present
author likes to remind the readers that by assuming the Beltrami
condition [3-4] to be valid in Eq. (2) then the second term in the
right-hand side of Eq. (2) will disappear (because $\nu \parallel
{\bf v}$; $\parallel$ means ' is parallel to'). There is no need
to presume $\eta'=0$ in Eq. (2)!
\newline In fact, Link solved
\begin{equation}
  \frac{\rho_s \kappa^2}{4\pi}\ln(\frac{1}{\xi k})
  \frac{\partial^2 {\bf u}}{\partial z^2} +\rho_s {\bf \kappa}
  \times (\frac{\partial {\bf u}}{\partial t}-{\bf v}_s)+
  {\bf  f}_0-\eta \frac{\partial {\bf u}}{\partial t}=0,
  \hspace*{12mm} \mbox{with}
\end{equation}
\begin{equation}
 {\bf f}_0 (z_i)=\sum_j (-1.70) F_m \frac{{\bf
 u}_j -{\bf r}_i}{r_p}\exp(-({\bf u}_j-{\bf r}_i)^2/2r_ p^2) ,
\end{equation}
where ${\bf u}$ is the displacement vector of the vortex w.r.t.
the $z$-axis  and  the nuclei are randomly placed at locations
${\bf r}_j$ in planes separated by $a$ and parallel to the $x$-$y$
plane (cf. the explanation for all notations above in [1]).
\newline Meanwhile the statement {\it By t=10$^3$, the vortex has
damped to a stationary pinned configuration with bends over a
characteristic length scale of $10 a$.} is of doubt since, once we
take a look at Fig. 3 of [1], we cannot  read out a value around
$10 a$ from only the $x$-$z$ plane.
%----------
As Link neglected quantum effects on the vortex motion and thermal
excitations (cf. page 2 of [1]) then it is natural to compare
Link's presentation with those based on the framework of the
semiclassical approximation (cf. [2]). Unfortunately we cannot
find out the contributions of vortex core [2,5-7] from [1]. This
lets us cast doubt about the applicable range of Link's
approach.\newline The other issue is about the presentation in [1]
(say, e.g., Figs. 2 and 3) extracted from the numerical
simulations : The vortex-free region [7] (relevant to the dynamics
of the vortex as well as the effective range of the vortex core)
is not identified in [1] and this will influence the calculated
results for the dynamics of a vortex [7]. To be specific,
considering an ideal fluid with the vorticity concentrated on a
smooth curve $\gamma$ [8], the vortex core should be regularized
by a value $\simeq$ (length($\gamma))^{-1/2}$ [8]. To keep the
circulation constant, its value changes as the product between the
core and the length of the curve ($\gamma$) is constant [8]. Not
to mention the stability issues and others relevant
[9-10].\newline The final remark is about the local deformation
induced by the moving vortex (cf. [7,10]) upon the (random)
lattice (consisting of $10^3$ nuclei per zone [1]). This issue was
neglected in [1] and the resulting effects were underestimated in
the numerical calculations  in [1]. This latter issue is also
crucial to the vortex pinning (w.r.t. the local lattice) site!

\end{document}